\documentstyle[proceedings,numreferences]{crckapb}
\makeindex
\begin{opening}
\title{THE INTERIOR OF BLACK HOLES AND THEIR ASTROPHYSICS}\\
\runningtitle{THE INTERIOR OF BLACK HOLES AND THEIR ASTROPHYSICS}
\author{I. V. ARTEMOVA}
\institute{Space Research Institute\\
Profsoyuznaya 84/32, Moscow 117997,Russia}
\author{I. D.NOVIKOV}
\institute{Theoretical Astrophysics Center\\
Juliane Maries Vej 30, DK-2100, Copenhagen, Denmark;\\
University Observatory\\
Juliane Maries Vej 30, DK-2100, Copenhagen,Denmark;\\
Astro-Space Center of Lebedev Physical Institute\\
Profsoyuznaya 84/32, Moscow 117997, Russia}
\end{opening}
\begin{document}
%\maketitle
\begin{abstract}

Gravity warps space and time into a funnel and generates a black hole when a
cosmic body undergoes a catastrophic collapse. What can one say about the interior
of a black hole? The important point is that inside a black hole the space radial
direction becomes time, and time becomes a space direction. The path into the 
gravitational abyss of the interior of a black hole is a progression in time. There 
is a peculiar region inside a black hole where some characteristics of the space-time
curvature become singular. We call this region {\it singularity}. The colossal tidal
 gravitational forces near singularity modify physical laws. Space and time are not 
only strongly curved near the singularity, but they split into quanta. The fall into 
the singularity is unstoppable for a body inside a black hole. This paper also 
addresses the following questions:

Can one see what happens inside a black hole?

Can a falling observer cross the singularity without being crushed?

Can new baby universes arise inside a black hole?

An answer to all these questions is probably ``yes''.
We give also a brief review of the modern black hole astrophysics.
\end{abstract}
\section{Introduction}

A black hole is perhaps the most fantastic of all conceptions of the human mind. 
Black holes are neither bodies nor radiation. They are clots of gravity. The study 
of black hole physics extends our knowledge of the fundamental properties of space 
and time. Quantum processes occur in the neighborhood of black holes, so that the 
most intricate structure of the physical vacuum is revealed. Even more powerful 
(catastrophically powerful) quantum processes take place inside black holes (in the 
vicinity of the singularity). One can say that black holes are a door to a new, 
very wide field of study of the physical world.
But probably the problems of the internal structure of black holes are a real great
challenge.

Inside a black hole the main sights are the singularity.

In this paper we want to outline the recent achievements in our understanding of the 
nature of the singularity (and beyond) inside a realistic, rotating black hole.

We give also a brief review of some problems of astrophysics of black holes.

For systematic discussion of the problems of the internal structure of black holes 
see [1-4]. Discussion of the problems of astrophysics of black holes see in [5-10].

\section{Interior of Black Holes}
 
The problem of black holes interior was the subject of a very active investigation 
last decades. There is a great progress in these researches. We know some important 
properties of the realistic black hole's interior, but some details and crucial 
problems are still the subject of much debate.

A very important point for understanding the problem of black hole's interior is the 
fact that the path into the gravitational abyss of the interior of a black hole is a 
progression in time. We recall that inside a spherical hole, for example, the radial 
coordinate is timelike. It means that the problem of the black hole interior is 
an {\it evolutionary problem}. In this sense it is completely different from a problem 
of an internal structure of other celestial bodies, stars for example, or planets.

In principle, if we know the conditions on the border of a black hole (on the event 
horizon), we can integrate the Einstein equations in time and learn the structure of 
the progressively deeper layers inside the black hole. Conceptually it looks 
simple, but there are two types of principal difficulties which prevent realizing 
this idea consistently.

The first difficulty is the following. The internal structure of a generic rotating 
black hole even soon after its formation depends crucially on the conditions on the 
event horizon at very distant future of the external observer (formally at the 
infinite future). This happens because the light-like signal can propagate from the 
very distant future to those regions inside a black hole which are deep enough in 
the hole. The limiting light-like signals which propagate from (formally) infinite 
future of the external observer form a border inside a black hole which is called 
a { \it Cauchy horizon}.

Thus, the structure of the regions inside a black hole depends crucially on the 
fate of the black hole at infinite future of an external observer. For example, 
it depends on the final state of the black hole quantum evaporation (because of the 
Hawking radiation), on possible collisions of the black hole with other black holes, 
or another bodies, and it depends on the fate of the Universe itself. It is clear 
that theoreticians feel themselves uncomfortable under such circumstances.

The second serious problem is related to the existence of a singularity inside a black 
hole. A number of rigorous theorems (see references in [1]) imply that singularities 
in the structure of spacetime develop inside black holes. Unfortunately these 
theorems tell us practically nothing about the locations and the nature of the 
singularities. It is widely believed today that in the singularity inside a realistic 
black hole the characteristics of the curvature of the spacetime tends to infinity. 
Close to the singularity, where the curvature of the spacetime approaches the Plank 
value, the Classical General Relativity is not applicable. We have no a final version 
of the quantum theory of gravity yet, thus any extension of the discussion of physics 
in this region would be highly speculative. Fortunately, as we shall see, these singular 
regions are deep enough in the black hole interior and they are { \it in the future } 
with respect to overlying and { \it preceding} layers of the black hole where curvatures 
are not so high and which can be described by well-established theory.

The first attempts to investigate the interior of a Schwarzschild black hole have 
been made in the late 70's. It has been demonstrated that in the absence of external 
perturbations, those regions of the black hole interior which are located long after 
the black hole formation are virtually free of perturbations, and therefore it can be 
described by the Schwarzschild geometry for the region with radius less than the 
gravitational radius. This happens because the gravitational radiation from aspherical 
initial excitations becomes infinitely diluted as it reaches these regions. But this 
result is not valid in general case when the angular momentum or the electric charge 
does not vanish. The reason for that is related to the fact that the topology of the 
interior of a rotating or/and charged black hole differs drastically from the 
Schwarzschild one. The key point is that the interior of this black hole
 possesses a { \it Cauchy horizon }. This 
is a surface of infinite blueshift. Infalling gravitational radiation propagates inside 
the black hole along paths approaching the generators of the Cauchy horizon, and the 
energy density of this radiation will suffer an infinite blueshift as it approaches the 
Cauchy horizon.

In general  the evolution with time into the black hole deeps looks like the following. 
There is a weak flux of gravitational radiation into a black hole through the horizon 
because of small perturbations outside of it. When this radiation approaches the Cauchy 
horizon it suffers an infinite blueshift. The infinitely blueshift radiation together 
with the radiation scattered by the curvature of spacetime inside the black hole results 
in a tremendous growth of the black hole internal mass parameter (``mass inflation'', 
after Poisson and Israel [11]) and finally leads to formation of the curvature 
singularity of the spacetime along the Cauchy horizon. The infinite tidal gravitational 
forces arise here. This result was confirmed by considering different models of the 
ingoing and outgoing fluxes in the interior of charged and rotating black holes. It was 
shown that the singularity at the Cauchy horizon is quite weak. In particular, the 
integral of the tidal force in the freely falling reference frame over the proper time 
remains finite. It means that the infalling object would then experience the finite 
tidal deformations which (for typical parameters) are even negligible. While an infinite 
force is extended, it acts only for a very short time. This singularity exists in a 
black hole at late times from the point of view of an external observer, 
but the singularity which arises just after the gravitational collapse of a 
star is much stronger. It seems likely that an observer falling into a black hole with 
the collapsing star encounters a crushing singularity, but an observer falling in a 
late times generally reaches a weak singularity.

\section{Quantum effects}

As we mentioned in Section 2 quantum effects play crucial role in the very vicinity of 
the singularity. In addition to that the quantum processes probably are important also 
for the whole structure of a black hole. Indeed, in the previous discussion we 
emphasized that the internal structure of black holes is a problem of evolution in time 
starting from boundary conditions on the event horizon for all moments of time up to 
the infinite future of the external observer.

It is very important to know the boundary conditions up to infinity because we observed 
that the essential events - mass inflation and singularity formation - happened along 
the Cauchy horizon which brought information from the infinite future of the external 
spacetime. However, even an isolated black hole in an asymptotically flat spacetime 
cannot exist forever.  It will evaporate by emitting Hawking quantum radiation. So far 
we discussed the problem without taking into account this ultimate fate of black holes.
Even without going into details it is clear that quantum 
evaporation of the black holes is crucial for the whole problem.

What can we say about the general picture of the black hole's interior accounting for 
quantum evaporation? To account for the latter process we have to change the boundary 
conditions on the event horizon as compared to the boundary conditions discussed above. 
Now they should include the flux of negative energy across the horizon, which is 
related to the quantum evaporation. The last stage of quantum evaporation, when the 
mass of the black hole becomes comparable to the Plank mass 
$m_{Pl}=(\hbar c/G)^{1/2}\approx 2.2\times 10^{-5} g$, is unknown. At this stage the 
spacetime curvature near the horizon reaches $l_{Pl}^{-2}$, where $l_{Pl}$ is 
the Plank length:

$l_{Pl}=\left(\frac{G\hbar}{c^{3}}\right)^{1/2}\approx 1.6\times 10^{-33} cm$.

This means that from the point of view of semiclassical physics a singularity arises 
here. Probably at this stage the black hole has the characteristics of an extreme black 
hole, when the external event horizon and internal Cauchy horizon coincide.

As for the processes inside a true singularity in the black hole's interior, they can 
be treated only in the framework of an unified quantum theory incorporating gravitation, 
which is unknown.

\section{Baby Universes inside a Black Hole?}

How the effects of quantum gravity could modify the structure of the spacetime 
singularity inside the black hole. To analyze this, let us consider a black hole which 
arises as a result of a spherically symmetric gravitational collapse. We know that the 
spacetime inside the black hole outside the collapsing matter can be described as an 
evolution of anisotropic homogeneous three-dimensional space. This metric has the 
Kasner-type asymptotic behavior near the singularity: the contraction of space in 
two directions is accompanied by expansion in the third direction. The curvature 
invariant 
$\it{R}^{2} = \it{R}_{\alpha \beta \gamma \delta} \it{R}^{\alpha \beta \gamma \delta}$
grows as $\it{R}^{2}$=$48M^{2}/r^{6}$, (we use units $c=1, G=1$).

Such behavior is a consequence of the classical equations which are valid until the 
spacetime curvature becomes comparable to the Plank one. Particle creation and vacuum 
polarization may change this regime. Quite general arguments allow one to suggest that 
the quantum effects may result in a decrease of the spacetime anisotropy, see 
references in [1]. Unfortunately, one cannot prove this result rigorously without 
knowing the physics at Plank scales.

Under these circumstances it is natural to use the following approach. One might assume 
that the notion of quantum average of a metric $g=<\hat{g}>$ is still valid in the 
regions under consideration, and the average metric $g$ obeys some effective equations.
We do not know these equations at the moment, but we might assume that these equations 
and their solutions obey some general properties and restrictions. For example, it is 
natural to require that the effective equations for $g$ in the low curvature limit 
reduce to the Einstein equations with possible higher-curvature corrections. It is also 
possible to assume that the future theory of quantum gravity would solve the problems of 
singularities of classical general relativity. One of the possibilities is that the 
equations of the complete theory would simply not allow dynamically infinite growth of 
the curvature, so that the effective  curvature $\it{R}$ of $g$ is bounded by the value 
of order $1/l^{2}$. This $\it{limiting}$ $\it{curvature}$ $\it{principle}$ was proposed 
by Markov [12], [13]. This principle excludes curvature singularity formation, so that 
the global properties of the solutions must change.

In the application to the problem of black hole interior the limiting curvature 
principle means that the singularity which, according to the classical theory exists 
inside a black hole, must be removed in the complete quantum theory. We cannot hope 
to derive this result without knowledge of the theory, but we may at least discuss 
and classify possibilities. Such approach is a natural first step, and it was used 
in a number of publications. We discuss here a singularity-free model of a black hole 
interior proposed by Frolov, Markov, and Mukhanov [14, 15]. According to this FMM-model, 
inside a black hole there exists a closed Universe instead of a singularity. It is 
assumed that the transition between Schwarzschild and de Sitter regimes is so fast that 
the transition region can be approximated by a thin spacelike shell. The presented 
model can be considered as ``the creation of a universe in the laboratory''. One of the 
assumptions of FMM and other similar models is that a ``phase transition'' to the de 
Sitter-like phase takes place at a homogeneous spacelike surface $r=r_{0}$. The 
presence of perturbations and quantum fluctuations, growing as $r \rightarrow 0$, 
could spoil the homogeneity. The bubbles of the new de Sitter-like phase could be 
formed independently at point separated by spacelike distances. For these reasons, 
one could expect that different parts of a black hole interior, can create spatially 
disconnected worlds.

\section{Can one see what happens inside a Black Hole?}

Is it possible for a distant observer to receive information about the interior of a 
black hole? Strictly speaking, this is forbidden by the very definition of a black hole. 
What we have in mind in asking this question is the following. Suppose there exists a 
stationary or static black hole. Can we, by using some device, get information about 
the region lying inside the apparent horizon?

Certainly it is possible if one is allowed to violate the weak energy condition. For 
example, if one sends into a black hole some amount of "matter" of negative mass, the 
surface of black hole shrinks, and some of the rays which previously were trapped 
inside the black hole would be able to leave it. If the decrease of the black hole mass 
during this process is small, then only a very narrow region lying directly inside the 
horizon of the former black hole becomes visible.

In order to be able to get information from regions not close the apparent horizon but 
deep inside an original black hole, one needs to change drastically the parameters of 
the black hole or even completely destroy it. A formal solution corresponding to such 
a destruction can be obtained if one considers a spherically symmetric collapse of 
negative mass into a black hole. The black hole destruction occurs when the negative 
mass of the collapsing matter becomes equal to the original mass of the black hole. 
In such a case an external observer can see some region close to the singularity. But 
even in this case the four-dimensional region of the black hole interior which becomes 
visible has a four-dimensional spacetime volume of order $M^{4}$. It is much smaller 
than four-volume of the black hole interior, which remains invisible and which is of 
order $M^{3}T$, where $T$ is the time interval between the black hole formation and 
its destruction (we assume $T\gg M$). The price paid for the possibility of seeing 
even this small part of the depths of the black hole is its complete destruction.

Does this mean that it is impossible to see what happens inside the apparent horizon 
without a destructive intervention? We show that such a possibility exists (Frolov and 
Novikov [16, 17]). In particular, we discuss a gedanken experiment which demonstrates 
that traversable wormholes (if only they exist) can be used to get information from 
the interior of a black hole practically without changing its gravitational field.
Namely, we assume that there exist a traversable wormhole, and its mouths are freely 
falling into a black hole. If one of the mouths crosses the gravitational radius 
earlier than the other, then rays passing through the first mouth can escape from the 
region lying inside the gravitational radius. Such rays would go through the wormhole 
and enter the outside region through the second mouth, see details in [1].

\section{Astrophysics of Black Holes}

The observational evidence at present is that black holes do exist in the Universe. 
Thus the singularities of spacetime which we discussed in the previous sections are 
real things.

A review of the black holes astrophysics we gave in [8]. Here we will give some 
additional remarks.

First of all we remind that modern astrophysics deals with two types of black holes 
in the universe.

(1) $\it{Stellar}$ $\it{black}$ $\it{holes}$, i.e. black holes of stellar masses that 
were born 
when massive stars died.

(2) $\it {Supermassive}$ $\it {black}$ $\it{holes}$ with masses up to $10^{9}M_{\odot}$
 and very recently (September 2002) some information appeared about possible discovery 
of the black holes of intermediate masses : ($4 \times 10^{3} - 2 \times 10^{4}$) solar 
masses in global cluster.

We start from the remarks concerning stellar black holes.

As we described in [8] very massive stars at the very end of their evolution probably 
turn into black holes. The evolution of stars in close binary systems differs from the 
evolution of solitary stars because of mass transfer from one star to another. The 
conclusions about masses of black hole progenitor in this case could be essentially 
different. In particular, a black hole can be produced in the binary where originally, 
besides a normal star, there was a neutron star. A black hole can be formed as a result
of the flux of matter from the star companion onto the neutron star, which finally makes 
the mass of the latter greater than the neutron mass limit (which is probably 
$M_{o} \approx 3M_{\odot}$).

Nowadays it is believed that Wolf-Rayet (WR) binaries may be progenitors of the binary 
systems with relativistic objects (neutron stars and black holes).

Wolf-Rayet stars are believed to be hot massive stars that are close to the end of 
their nuclear burning phase. Formation of neutron stars and black holes in close binary 
systems is going through the WR stage of initially more massive stars in the binary 
systems.

Recently Cherepashchuk [6] has analyzed the observed properties of Wolf-Rayet stars and 
relativistic objects in close binary systems. He calculated the final masses 
$M_{CO}^{f}$ for the carbon-oxygen cores of WR + O binaries taking into account the 
radial loss of matter via stellar wind, which depends on the mass of star. The analysis 
includes new data on the clumpy structure of WR winds, which appreciably decreases the 
required mass-loss rates $\dot{M}_{WR}$ for the WR stars, (see Fig.1, top and bottom 
panels). The masses $M_{CO}^{f}$ lie in the range $(1-2)M_{\odot}-(20-44)M_{\odot}$
and have a continuous distribution. The masses of the relativistic objects $M_{x}$ 
are $1-20M_{\odot}$ and have a bimodal distribution: the mean masses for neutron stars 
and black holes are $1.35 \pm 0.15 M_{\odot}$ and $8-10M_{\odot}$ respectively, with a gap 
from $2-4M_{\odot}$ in which no neutron stars and black holes are observed in close 
binaries, (see Fig.1, middle panel). The mean final CO-core mass is 
$\overline{M}_{CO}^{f}=7.4-10.3M_{\odot}$, close to the mean mass for the black holes. 
This suggests that it is not only the mass of the progenitor that determines the nature 
of the relativistic object, but other parameters as well - rotation, magnetic field, etc.

Now some remarks concerning supermassive black holes. 

At the beginning we will remind some facts.

About one percent of all galactic nuclei eject radio-emitting plasma and gas clouds, 
and are themselves powerful sources of radiation in the radio, infrared, and 
especially, the ``hard'' (soft wavelength) ultraviolet, X-ray and gamma regions of the 
spectrum. The full luminosity of the nucleus reaches in some cases 
$L \approx 10^{47} erg/s$. This is millions of times greater than the luminosity of 
the nuclei of more quite galaxies, such as ours. These objects were called 
$\it {active}$ $\it {galactic}$ $\it {nuclei}$ (AGN). Practically all the energy of 
activity and of the giant jets released by galaxies originates from the centers of 
their nuclei.

Quasars form a special subclass of AGN. Their characteristic property is that their 
total energy release is hundreds of times greater than the combined radiation of all 
the stars in a large galaxy. At the same time the average linear dimensions of the 
radiating regions are small: a mere one-hundred-millionth of the linear size of a 
galaxy. Quasars are the most powerful energy sources registered in the Universe to 
date. What processes are responsible for the extraordinary outbursts of energy from 
AGN and quasars?

At the beginning of 1980s a paradigm was established in which AGN are powered by black 
holes accreting gas stars [18].

\begin{figure}
\vspace{13cm} % amount of vertical space needed
\caption{Distribution of the final masses $M_{CO}^{f}$ for two different versions of 
the theory (see details in [6]), top and bottom panels; and distribution of the
masses of the relativistic objects $M_{x}$, middle panel. According to [6].}
\end{figure}

Clearly, the processes taking place in quasars and other galactic nuclei are still a 
mystery in many respects. But the suggestion that we are witnessing the work of a 
supermassive black hole with an accretion disk seems rather plausible. Rees (1990) 
advocates a hypothesis that the massive black holes are not only in the active galactic 
nuclei but, also in the centers of ``normal'' galaxies (including nearby galaxies and our 
own Milky Way). They are quiescent because they now starved of fuel (gas). Observations 
show that galactic nuclei were more active in the past. Thus, ``dead quasars'' (massive 
black holes without fuel) should be common at the present epoch.

 How can this black holes be detected? It has been pointed out that holes produce 
cusp-like gravitational potentials and hence they should produce cuspy-like density 
distributions of the stars in the central regions of galaxies. Some authors have 
argued that the brightness profiles of the central regions of particular galaxies imply 
that they contain black holes. However the arguments based only on surface brightness 
profiles are inconclusive. The point is that a high central number density of stars in 
a core with small radius can be the consequence of dissipation, and a cusp-like profile 
can be the result of anisotropy of the velocity dispersion of stars. Thus these 
properties taken alone are not sufficient evidence for the presence of a black hole.

The reliable way to detect black holes in the galactic nuclei is analogous to the case 
of black holes in binaries. Namely, one must prove that there is a large dark mass in a 
small volume, and that it can be nothing but a black hole. In order to obtain such a 
proof we can use arguments based on both stellar kinematics and surface photometry of 
the galactic nuclei.

If the distribution of the mass $M$ and the luminosity $L$ as functions of the radius are 
known we can determine the mass-to-light ratio $M/L$ (in solar units) as a function of 
radius. This ratio is well known for different types of stellar populations. As a rule 
this ratio is between 1 and 10 for elliptical galaxies and globular cluster 
(old stellar population dominates there). If for some galaxy the ratio $M/L$ is almost 
constant at rather large radii (and has a ``normal'' value between 1 and 10) but rises 
rapidly toward values much larger than 10 as one approaches the galactic center, then 
this is the evidence for a central dark object (probably a black hole).

The technique described above has been used to search for black holes in galactic 
nuclei. Another possibility is to observe rotational velocities of gas in the vicinity 
of the galactic center.

The modern search techniques and results for the ground-based detections see in [19] 
and [20]. 

Here we discuss the results from Hubble Space Telescope (Kormendy and Gebhardt [7]).
According to [7], dynamical black hole (BH) detections are available for at least 37 
galaxies (see Table 1). Their conclusions are the following.

(1) BH mass 
correlates with the luminosity of the bulge component of the host galaxy, albeit with 
considerable scatter. The median BH mass fraction is $0.13 \%$ of the mass of the bulge.

(2) BH mass correlates with the mean velocity dispersion of the bulge inside its 
effective radius, i.e., with how strongly the bulge stars are gravitationally bound 
to each other. For the best mass determinations, the scatter is consistent with the 
measurement errors.

(3) BH mass correlates with luminosity of the high-density central component in disk 
galaxies independent of whether this is a real bulge (a mini-elliptical, believed to 
form via a merger-induced dissipative collapse and starburst) or a ``pseudobulge'' 
(believed to form by inward transport of disk material).

(4) BH mass does not correlate with the luminosity of galaxy disks. If pure disks 
contain BHs (and active nuclei imply that some do), then their masses are much smaller 
than $0.13 \%$ of the mass of the disk.

We conclude that present observations show no dependence of BH mass on the details of 
whether BH feeding happens rapidly during a collapse or slowly via secular evolution 
of the disk. The above results increasingly support the hypothesis that the major 
events that form a bulge or elliptical galaxy and the main growth phases of its 
BH - when it shone like a quasar - were the same events.

In the future we expect the most fundamental progress in the work on the black holes 
astrophysics from gravitational wave astronomy.

%\begin{table}[htb] % the table will be place at the separate page
\begin{table}[ht]
\begin{center}
\caption{Census of Supermassive Black Holes (2001 March), according to [7].}
\begin {tabular}{lllllll}
\hline
Galaxy &Type&$M_{B,bulge}$&$M_{BH}$$(M_{low},M_{high})$&$\sigma_{e}$& D &$r_{cusp}$\\
       &    &             &         $M_{\odot}$        &  (km/sec)  &(Mpc)&(arcsec) \\
\hline
Galaxy & Sbc & -17.65 & 2.6 (2.4 - 2.8) e6 & 75  & 0.008 & 51.40\\
M31    & Sb   & -19.00 & 4.5 (2.0 - 8.5) e7 & 160 & 0.76 & 2.06\\
M32    & E2   & -15.83 & 3.9 (3.1 - 4.7) e6 & 75  & 0.81 & 0.76\\
M81    & Sb   & -18.16 & 6.8 (5.5 - 7.5) e7 & 143 & 3.9  & 0.76\\
NGC 821 & E4   & -20.41 & 3.9 (2.4 - 5.6) e7 & 209 & 24.1 &0.03\\
NGC 1023 & S0  & -18.40 & 4.4 (3.8 - 5.0) e7 & 205 & 11.4 &0.08\\
NGC 2778 & E2  & -18.59 & 1.3 (0.5 - 2.9) e7 & 175 & 22.9 &0.02\\
NGC 3115 & S0  & -20.21 & 1.0 (0.4 - 2.0) e9 & 230 & 9.7  &1.73\\
NGC 3377 & E5  & -19.05 &1.1 (0.6 - 2.5) e8 & 145 & 11.2 &0.42\\
NGC 3379 & E1   & -19.94 & 1.0 (0.5 - 1.6) e8 & 206 & 10.6 &0.20\\
NGC 3384 & S0   & -18.99 & 1.4 (1.0 - 1.9) e7 & 143 & 11.6 &0.05\\
NGC 3608 & E2   & -19.86 & 1.1 (0.8 - 2.5) e8 & 182 & 23.0 &0.13\\
NGC 4291 & E2  & -19.63 & 1.9 (0.8 - 3.2) e8 & 242 & 26.2 &0.11\\
NGC 4342 & S0     &-17.04 & 3.0 (2.0 - 4.7) e8 & 225 & 15.3 &0.34\\
NGC 4473 & E5       & -19.89 & 0.8 (0.4 - 1.8) e8 & 190 & 15.7 &0.13\\
NGC 4486B & E1      &-16.77 & 5.0 (0.2 - 9.9) e8 & 185 & 16.1 &0.81\\
NGC 4564 & E3     &-18.92 & 5.7 (4.0 - 7.0) e7 & 162 & 15.0 &0.13\\
NGC 4594 & Sa   &-21.35 & 1.0 (0.3 - 2.0) e9 & 240 & 9.8 &1.58\\
NGC 4649 & E1  &-21.30 & 2.0 (1.0 - 2.5) e9 & 375 & 16.8 &0.75\\
NGC 4697 & E4   &-20.24 & 1.7 (1.4 - 1.9) e8 & 177 & 11.7 &0.41\\
NGC 4742 & E4  &-18.94 & 1.4 (0.9 - 1.8) e7 & 90 & 15.5 &0.10\\
NGC 5845 & E  &-18.72 & 2.9 (0.2 - 4.6) e8 & 234 & 25.9 &0.18\\
NGC 7457 & S0  &-17.69 & 3.6 (2.5 - 4.5) e6 & 67 & 13.2 &0.05\\
\hline
NGC 2787 & SB0 & -17.28 & 4.1 (3.6 - 4.5) e7 & 185 & 7.5 &0.14\\
NGC 3245 & S0 & -19.65 & 2.1 (1.6 - 2.6) e8 & 205 & 20.9 &0.21\\
NGC 4261 & E2 & -21.09 & 5.2 (4.1 - 6.2) e8 & 315 & 31.6 &0.15\\
NGC 4374 & E1 & -21.36 & 4.3 (2.6 - 7.5) e8 & 296 & 18.4 &0.24\\
NGC 4459 & SA0 & -19.15 & 7.0 (5.7 - 8.3) e7 & 167 & 16.1 &0.14\\
M87      & E0  & -21.53 & 3.0 (2.0 - 4.0) e9 & 375 & 16.1 &1.18\\
NGC 4596 & SB0 & -19.48 & 0.8 (0.5 - 1.2) e8 & 136 & 16.8 &0.22\\
NGC 5128 & S0 & -20.80 & 2.4 (0.7 - 6.0) e8 & 150 & 4.2 &2.26\\
NGC 6251 & E2 & -21.81 & 6.0 (2.0 - 8.0) e8 & 290 & 106 &0.06\\
NGC 7052 & E4 & -21.31 & 3.3 (2.0 - 5.6) e8 & 266 & 58.7 &0.07\\
IC 1459  & E3 & -21.39 & 2.0 (1.2 - 5.7) e8 & 323 & 29.2 &0.06\\
\hline
NGC 1068 & Sb & -18.82 & 1.7 (1.0 - 3.0) e7 & 151 & 15 & 0.04\\
NGC 4258 & Sbc & -17.19 & 4.0 (3.9 - 4.1) e7 & 120 & 7.2 & 0.36\\
NGC 4945 & Scd & -15.14 & 1.4 (0.9 - 2.1) e6 & & 3.7 &     \\
\hline
\end{tabular}
\end{center}
{Notes - BH detections are based on stellar dynamics (top group),  
ionazed gas dynamics (middle) and maser dynamics (bottom). 
Column 3 is the B-band absolute magnitude of the bulge part of the 
galaxy. Column 4 is the BH mass $M_{BH}$ with error bars 
$(M_{low}, M_{high})$. Column 5 is the galaxy's velocity dispersion 
(see Figure 1). Column 6 is the distance (Tonry $\it {et}$ $\it{al.}$
 2001). Column 7 is the radius of the sphere of influenece of the BH.}
%\end{center}
\end{table}

\section{Probing black holes with gravitational waves}

Observations of stellar and massive black holes in optics and X-and $\gamma$-rays do 
not provide us with direct information about spacetime regions close to a black hole, 
since the radiation is generated in regions far from horizon. To explore the 
region close to the horizon in detail may well require using a new 
information channel in astrophysics - gravitational waves. With the 
construction of new gravitational wave observatories this option becomes 
very important.

Among the most promising sources of gravitational waves which can be observed 
by the gravity wave detectors are astrophysical compact binaries. Three types 
of compact binaries are mainly discussed: neutron star - neutron star (NS/NS) 
binaries, neutron star - black hole (NS/BH) binaries, and black hole - black 
hole (BH/BH) binaries. Because of the emission of gravitational waves at some 
stage of their evolution, compact binaries enter the inspiral phase which ends 
with a coalescence. During these final stages of the binary system evolution 
they emit powerfull gravitational waves.

An international network of ground-based gravitational wave detectors is now 
under construction. It includes two detectors of the American Laser 
Interferometer Gravitational-wave Observatory (LIGO) [21], the French/Italian 
3-kilometer-long arms interferometer VIRGO near Pisa (Italy) [22], and the 
British/German 600-meter interferometer GEO-600 near Hannover (Germany) [23].

The LIGO detector, which is now under construction, consists of two vacuum 
facilities with two 4-km-long orthogonal arms. One of these detectors is in 
Hanford (state Washington) and the other in Livingston (state Louisiana). 
Their coincident operation will start in 2002. Gravitational waves coming 
from far-distant sources effectively change the relative length of the arms, 
which can be measured by the phase shift between two lasers beams in the two 
orthogonal arms. With expexted accuracies of the 
arm-length difference $\Delta L\sim 10^{-16}$ cm, the expected 
sensitivity of the detector would be 
$\Delta L/l \sim 10^{-21} - 10^{-22}$. This sensitivity will be achieved 
in LIGO within the frequency range from 40 to 120 Hz. The efficiency of 
LIGO is effectively reduced by photon counting statistic (`shot noise') 
at higher frequency and by seismic noise at lower frequency. The LIGO 
facilities are designed to house many successive generations of upgraded 
interferometers. The second generation, LIGO II, is planned to start to 
be designed in 2005, and to be observing before 2007. Working in the same 
frequency range, it will have an approximately two orders of magnitude 
higher sensitivity. Table 2 gives the limiting distances up to which the 
LIGO detectors would be able to observe different types of binaries.

%\begin{table}[htb] % the table will be place at the separate page
\begin{table}[ht]
\begin{center}
\caption{List of the sources detectable by LIGO I and LIGO II. 
Neutron stars are assumed to have mass $1.4M\odot$, and black holes are 
assumed to have mass $10M\odot$. Data from Ref.[21].}
\begin {tabular}{lll}
\hline
Systems & Distance for LIGO I & Distance for LIGO II\\
\hline
Inspiral NS/NS binaries & 20 Mpc & 450 Mpc\\
Inspiral NS/BH binaries & 40 Mpc & 1000 Mpc\\
Inspiral BH/BH binaries & 100 Mpc & 2000 Mpc\\
\hline
\end{tabular}
\end{center}
\end{table}

Black hole binariy evolution and its emitted gravitaional waveforms can 
be divided into the following three stages: inspiral, coaliscence, and 
ringdown. The inspiral epoch for a BH/BH binary requires post-Newtonian 
expansions for its understanding and is qualitatively the same as for 
other compact binaries. Gravitational radiation during coalescence and 
ringdown epochs contains information which allows the BH/BH case to be 
singled out. Supercomputer simulations are required to determine the 
dynamics of two merging black holes and to produce templates which can 
be used to decode the information encoded in emitted gravitational waves. 
The ringdown epoch is much better understood. At this stage, two initial 
black holes form a new final one, which is in a very excited state. Its 
further evolution involves a decay of these excitations. These 
excitations are a nonlinear superposition of quasi-normal modes. The decay 
of the quasi-normal modes produces a characteristic `ringing' in the 
gravitational waveforms.

Gravitational waves emitted at the stages of BH/BH coalescence and 
ringdown carry information about the highly nonlinear, large-amplitude 
dynamics of spacetime curvature, and for this reason the study of these 
signals tests Einstein gravitational equations in their full complexity. 
Table 3 gives an estimate of the amplitude signal-to-noise (S/N) ratio 
for coalescences at a 1000-Mpc distance for two black holes of equal 
mass.

%\begin{table}[htb] % the table will be place at the separate page
\begin{table}[ht]
\begin{center}
\caption{Amplitude signal-to-noise (S/N) ratio for coalescences at 
1000-Mpc distance for two black holes of equal mass. Data from Ref. [21].}
\begin {tabular}{lll}
\hline
BH/BH coalescences & S/N for LIGO I & S/N for LIGO II\\
\hline
$10M_{\odot}/10M_{\odot}$ & 0.5 & 10\\
$25M_{\odot}/25M_{\odot}$ & 2 & 30\\
$100M_{\odot}/100M_{\odot}$ & 4 & 90\\
\hline
\end{tabular}
\end{center}
\end{table}

The time- and length-scales for double black-hole dynamics (including the 
gravitational radiation from such systems) are proportional to the total 
mass. Other parameters (such as the black hole mass ratio, black hole 
angular momentum and so on) enter through dimensionless combinations. The 
total number of cycles spent in the LIGO/VIRGO band for a BH/BH of 
$10M_{\odot}$ is about 600. These detectors will be able to detect and study 
gravitational waves emitted during last few minutes of their evolution for 
black hole binaries with a total mass of up to $10^{3}M_{\odot}$. For larger 
masses, a gravitational wave detector must have a much lower frequency band. 
Future space-based gravitational wave interferometers will work in this band. 
LISA is an example of such a project.

The Laser Interferometer Space Antenna (LISA) consists of 3 spacecraft 
flying $5 \times 10^{6}$ km apart in the shape of an equilateral triangle. 
The center of the triangle will be at the ecliptic plane at the same 
distance from the Sun as the Earth and $20^{o}$ behind the Earth on the 
orbit. The three spacecraft will act as a giant interferometer measuring 
distortions in space caused by gravitational waves. This project was 
proposed in 1993 by the United States and European scientists as a joint 
NASA/ESA (National Aeronautics and Space Administration/European Space 
Agency) mission. If approved, the project will start in 2005 with a 
launch planned around 2010 [24].

The frequency band of LISA covers $10^{-4} - 1 Hz$, that is 10,000 times 
lower than the frequency band of LIGO/VIRGO. Its sensititvity in this 
frequency band will be at the level of $10^{-23}$. LISA will be able to 
register gravitational waves emitted by BH/BH binaries for a total mass 
in the range $10^{3}M_{\odot} - 10^{8}M_{\odot}$ (massive and suremassive 
black holes), away from each other by a distance corresponding to redshifts 
of $z \sim 3000$. Since it is very unlikely that massive and supermassive
 black holes form so early (until they are primordial), this means that 
LISA will be able to observe pratically $\it {all}$ coalescing black hole 
binaries in the visible universe within this range of mass.

For a discussion of gravitational-wave radiation form colliding black 
holes it is very important to know how many BH/BH binaries exist in 
the universe. Unfortunately, this is not known. The scatter between 
the most optimistic and most pessimistic estimates is quite wide. 
However, for BH/BH binaries with a total mass of $5-50M_{\odot}$ that 
are created from main-sequence progenitors, one can expect a coalescence 
rate in our Galaxy of 1 per 1-30 million years [25-27]. If these 
estimates are correct, LIGO/VIRGO will see one coalescence per year 
for such binaries up to the distance of 300-900 Mpc. The event rate 
for supermassive black hole coalescences is much more 
uncertain - from 0.1 to 1000 per year. But even for the pessimistic 
rate value, LISA will be able to observe 3 BH/BH binaries with a total 
mass of $3,000 -10^{5}M_{\odot}$ that are 30 years away from their 
final coalescence [27,28]. For all details see [29].

To summarize, there is a good chance that in the near future 
gravitational waves from coalescing black holes will be observed and, 
hence, for the first time we shall be able to probe almost directly 
our theoretical predictions concerning black holes.

\vspace{0.5cm}

{\bf Acknowledgments}

This paper was supported in part by the Russian Foundation for Basic 
Research 00-02-16135, in part by the Danish natural Science Research 
Council through grant No. 9701841 and also in part by Danmarks 
Grundforskningsfond through its support for establishment of the TAC.

\index{first}
\end{document}